\documentclass{article}

\def\bfg{\hbox{\bf g}}
\def\bfe{\hbox{\bf e}}
\def\bfone{\hbox{\bf 1}}
\def\Tr{\mathop{\rm Tr}\nolimits}

\def\half{{\textstyle {1\over2}}}
\def\diag{\mathop{\rm diag}\nolimits}

\def\pmb#1{\setbox0=\hbox{$#1$}%
  \kern-.025em\copy0\kern-\wd0
  \kern.05em\copy0\kern-\wd0
  \kern-.025em\raise.0433em\box0}
\def\pmbs#1{\setbox0=\hbox{$\scriptstyle #1$}%
  \kern-.0175em\copy0\kern-\wd0
  \kern.035em\copy0\kern-\wd0
  \kern-.0175em\raise.0303em\box0}

    \def\bfsbeta{\pmbs{\beta}}

\makeatletter    
     
\def\eqalign#1{\null\,\vcenter{\openup\jot\m@th \let\\=\crcr 
  \ialign{\strut\hfil$\displaystyle{##}$&$\displaystyle{{}##}$\hfil
      \crcr#1\crcr}}\,}
\makeatother

\def\journalfont{\it}         
\def\jou#1{{\journalfont #1\ }}
\def\joudef#1#2{\def #1{\jou{\ignorespaces #2}}}

\joudef{\aaa}{  Astron.\ Astrophys.}
\joudef{\aip}{  Adv.\ Phys.}
\joudef{\am}{   Ann.\ Math.}
\joudef{\ap}{   Ann.\ Phys.\ (N.Y.)}
\joudef{\aop}{   Ann.\ Phys.\ (N.Y.)}
\joudef{\apj}{  Astrophys.\ J.}
\joudef{\cjp}{  Can.\ J.\ Phys.}
\joudef{\cmp}{  Commun.\ Math.\ Phys.}
\joudef{\cqg}{  Class.\ Quantum Grav.}
\joudef{\grg}{  Gen.\ Relativ.\ Grav.}
\joudef{\ijtp}{ Int.\  J.\ Theor.\ Phys.}
\joudef{\jmp}{  J.\ Math.\ Phys.}
\joudef{\jpamg}{ J.\ Phys.\ A: Math.\ Gen.}
\joudef{\mnras}{ Mon.\ Not.\ R.\ Ast.\ Soc.}
\joudef{\nat}{  Nature}
\joudef{\ncim}{ Nuovo Cim.}
\joudef{\nucp}{ Nuc.\ Phys.}
\joudef{\ncb}{  Il Nuovo Cimento ``B}
\joudef{\pl}{   Phys.\ Lett.}
\joudef{\pr}{   Phys.\ Rev.}
\joudef{\prep}{ Phys.\ Rep.}
\joudef{\prl}{  Phys.\ Rev.\ Lett.}
\joudef{\ptp}{  Prog.\ Theor.\ Phys.}
\joudef\rmp{  Rev.\ Mod.\ Phys.}
\joudef\spj{  Sov.\ Phys.\ JETP}

\def\C#1#2#3{C^{#1}{}_{#2#3}}

\def\gm{\hbox{\rm g}}

\def\bfE{\hbox{\bf E}}
\def\bfR{\hbox{\bf R}}

\def\alphab{\overline{\alpha}\null}
\def\betab{\overline{\beta}\null}

\def\dt{\,\dot{}\,\,}
\def\ddt{\,\ddot{}\,\,}

\def\nthree{n^{(3)2}}
\def\none{n^{(3)}n^{(1)}}

\def\source{^{\rm source}}
\def\dee{_{\rm D}}

\def\ttg{_{\rm TTG}}
\def\gee{^{\rm G}}
\def\adm{\rm ADM}

\begin{document}

\title{Exact Cosmological Solutions \\
of Gravitational Theories}
\author{Robert T. JANTZEN\\
Villanova University, Villanova, PA 19085, USA and\\
International Center for Relativistic Astrophysics,\\
Department of Physics, University of Rome, I-00185 Rome, Italy}
\date{Received 22 September 1986\\ 
Physics Letters B186, 290--296 (1987)}

\maketitle

\begin{abstract}
A global picture is drawn tying together most exact cosmological solutions
of gravitational theories in four or more spacetime dimensions.
\end{abstract}

\section{Introduction}

The search for exact solutions of gravitational
theories in four spacetime dimensions has often become an end in itself,
a game in which one seeks to write down closed form expressions or at least
quadratures which solve
{\it some} particular case of {\it some} set of equations,
without worrying about interpretation, placing them into context or drawing
some significant conclusion about properties of the original equations.
Now this game, increasingly more difficult to play in four dimensions, is 
escaping
into higher dimensions where there is little resistance.  
This article wishes to point out that almost all known exact cosmological
solutions of deterministic field equations can be understood in terms of
a few simple ideas. (Perfect fluids with no equation of state or imperfect
fluids are meant to be excluded by the phrase deterministic field equations.)
The very existence of these exact solutions and their relationship to each
other depends crucially on the underlying structure of the field equations
themselves, a fact which is rarely appreciated.

These remarks are confined to spatially homogeneous
classical solutions of gravitational theories
in which the purely gravitational variables have Einsteinlike equations,
i.e., do not involve ``higher derivatives". For such theories a well developed
Lagrangian/Hamil\-ton\-ian 
formulation of the field equations helps to understand
the exact solutions in terms of an elegant geometrical picture.  This approach,
initiated by Arnowit, Deser and Misner [1] for  Einstein's equations in four
dimensions,  was developed by Misner [2--4]
as a powerful way of viewing the dynamics
of spatially homogeneous cosmological models.
It extends to higher dimensional theories in a natural way, including 
Brans-Dicke and supersymmetric variations of Einstein's theory coupled to
various matter sources. Henneaux [5] gives the details for ten-dimensional
supergravity, for example.

\section{Spatially Homogeneous Models}

By choosing the time coordinate
lines orthogonal to the spatial sections, the $d$-dimensional
spacetime metric  can be
expressed at least locally in the form [6--9]
$$ {}^{(d)}\gm =- N(t)^2\, dt\otimes dt + g_{ab}(t)\omega^a \otimes \omega^b
\eqno(1)$$
on the manifold $R\times (G/H)$, where $\{\omega^a\}$ are
1-forms on $G/H$ dual to the frame $\{e_a\}_{a=1,\ldots,D}$ (let $D=d-1$)
characterized by
structure functions $\C abc=\omega^a([e_a,e_b])$ on $G/H$.  These are
constants in the simply transitive case where $H$ is the identity subgroup of
the isometry group 
$G$ and $\{e_a\}$ is a left invariant
frame on $G/H=G$
(i.e., a basis of the Lie algebra of $G$), but they are functions on
the left coset space
$G/H$ in the nontrivial multiply transitive case, where $H$ is a continuous
subgroup.
In the first case the
matrix $\bfg\equiv(g_{ab})$ is an arbitrary positive definite symmetric matrix,
while in the second case it must satisfy additional linear 
constraints [8,9].
Additional spatial symmetries in the first case lead to similar constraints;
such cases might be called trivially multiply transitive as opposed to those
``nontrivial"
models which have no subgroup acting simply transitively on the orbits.

The choice of lapse function $N(t)$ determines the parametrization of the
family of homogeneous spatial sections, i.e., the time function.
The proper time $\tau$, defined by
$d\tau =N(t)dt=\omega^{\bot} $ corresponds to unit lapse.  In expanding
models with an initial singularity, $\tau$ is usually chosen so that
this singularity occurs at $\tau=0$.  Defining $\omega^0=dt$ leads to the
spacetime 1-forms $\omega^{\alpha}$ dual to the frame $\{e_{\alpha}\}_{\alpha
=0,1,\ldots,D}$, where $e_0=\partial/\partial t$ and $e_{\bot}=N^{-1}e_0$
is the unit normal to the spatial sections.
Clearly the choice of time function should be made to simplify the equations
of motion for $\bfg= (g_{ab})$.  In order to obtain exact solutions
(at least up to quadratures) one must decouple the equations in some way,
a problem which depends critically on the choice of time.

For the sake of brevity, consider only models which are ``diagonalizable" and
therefore can be assumed to be diagonal, i.e., the matrix $\bfg$
is diagonal
$$\bfg=\bfg\dee =\diag(g_{11},\ldots,g_{DD})
\eqno(2)$$
and hence its natural logarithm is an arbitrary diagonal matrix which can be 
decomposed into its pure trace and tracefree parts
$$\eqalign{ &\bfg\dee =e^{2\alpha} e^{2\bfsbeta}\ ,\cr
 &\half \ln \bfg\dee = \alpha \bfone +\beta^A\bfe_A\ ,\cr
 &\Tr \bfe_A=0\ ,\qquad \Tr \bfe_A \bfe_B = D(D-1)\delta_{AB}\ ,\cr}
\eqno(3)$$
where the normalization of the tracefree basis
is the most natural one in view of considerations described below.
Most known exact solutions fall into this diagonal
class. The existence of diagonal
solutions of the field equations depends on the symmetry type, the choice of
frame and the form of the field equations (including the symmetry of the source,
etc.) and is usually connected in some way with discrete or continous
spacetime symmetry of some kind.

Even though the present remarks will be confined to such diagonal solutions,
a similar analysis of the few exact solutions not of this form is even more
crucial due to their complexity.  Moreover, it is absolutely essential to
rely on these ideas to
appreciate the general spatially homogeneous case, which does not admit
exact solution [10,11].

\section{Field Equations}

Since we wish to focus on the gravitational properties of higher dimensional
models, we consider the Einstein equations with all other fields lumped into
the energy-momentum tensor (including a possible cosmological constant).
This assumes that the gravitational part of the
classical field equations can be represented in this way, thus excluding
``higher derivative" theories.  These equations
$$ 0=M_{\alpha \beta}\equiv |{}^{(d)}g|^{1/2} ({}^{(d)}G_{\alpha \beta}-
  k T_{\alpha\beta})\ , \qquad |{}^{(d)}g|^{1/2}  = Ng^{1/2} 
\eqno(4)$$
(where metric quantities without the leading superscript `${}^{(d)}$' refer
to the spatial metric) naturally split into evolution equations $0=M_{ab}$
which evolve the spatial metric and constraints on the solutions of
those equations
$$ \eqalign{
&0= \mathcal{H}\equiv 2N^{-1}M^{\bot}{}_{\bot} \equiv 2M^0{}_{\bot} \qquad
(\hbox{super-Hamiltonian constraint})\ ,\cr
&0= \mathcal{H}_a\equiv 2N^{-1}M^{\bot}{}_a \equiv 2M^0{}_a \qquad
(\hbox{supermomentum constraint})\ .\cr}
\eqno(5)$$

The Einstein equations may also be written in ``Ricci form"
$$ 0=P_{\alpha\beta}\equiv {}^{(d)}R_{\alpha\beta} - k E_{\alpha\beta}\ ,
\qquad E_{\alpha\beta} \equiv T_{\alpha\beta}-(d-2)^{-1}T^{\gamma}{}_{\gamma}
\,g_{\alpha\beta}\ ,
\eqno(6)$$
a form very convenient for static solutions in discussions of spontaneous
compactification.  However, for dynamical solutions a compromise is preferable,
namely the Ricci evolution equations $0=P_{ab}$, supplemented by the Einstein
constraints rather than more equations involving second time derivatives.
For our purposes it is crucial to choose the lapse function to be an
explicit function of the spatial metric given by
$$ N\ttg=2D(D-1) g^{1/2}\ ;
\eqno(7)$$
this choice of time will be called Taub time gauge. In fact, the Ricci
evolution equations are just the variational equations obtained from the
usual ADM Einstein Lagrangian [12]
with this explicit choice of the lapse, although for some symmetry types
one must add a nonpotential force to the system to get the correct equations.
For diagonal models, these equations are explicitly
$$ (\ln \bfg\dee)\,\ddot{}\,\, =-2N\ttg^2(\bfR-k\bfE)\ ,
\eqno(8)$$
where $\bfR=(R^a{}_b)$ is the matrix of mixed components of the spatial Ricci
tensor and $\bfE=(E^a{}_b)$.

In general, the evolution equations may be obtained from the ADM Lagrangian
or its associated Hamiltonian
$$\eqalign{ &L_{\adm}= T-U\ , \qquad H_{\adm}=T+U=N\mathcal{H}\ ,\cr
 & T=(4N)^{-1}\mathcal{G}^{abcd}\dot{g}_{ab}\dot{g}_{cd}\ ,\cr
 & U= U\gee +U\source\ ,\cr
 & U\gee=N(-g^{1/2}R)\ , \qquad U\source= N(-2 g^{1/2} k T^{\bot}{}_{\bot})\ ,
\cr}
\eqno(9)$$
supplemented by a nonpotential force [6] defined by
$$ -d(N^{-1}U) + Q^{ab}dg_{ab}= -g^{1/2}(G^{ab}-kT^{ab})dg_{ab}
\eqno(10)$$
for some symmetry types, where the exterior derivative is the one on the
space $\mathcal{M}$ of metric matrices, on which the DeWitt [13] metric is a
Lorentz metric
$$ 2N^{-1}\mathcal{G} =2N^{-1}g^{1/2} (g^{a(c}g^{d)b} - g^{ab}g^{cd})dg_{ab}\otimes
dg_{cd}\ .
\eqno(11)$$
Allowing the lapse to have a factor depending explicitly on the spatial metric
is equivalent to conformally rescaling this metric by that factor [4].
In particular, the Taub time gauge removes the metric determinant factor
and makes $\{\alpha,\beta^A\}$ natural Lorentz orthonormal coordinates
on the flat diagonal submanifold $\mathcal{M}\dee$ with respect to the rescaled
DeWitt metric (a conventional factor of 2 is missing from the kinetic energy)
$$ 2N\ttg^{-1}\mathcal{G}|_{\mathcal{M}\dee}
 = -d\alpha\otimes d\alpha + \delta_{AB}
d\beta^A \otimes d\beta^B\ .
\eqno(12)$$
$\alpha$ is the natural time variable on the space, and $\beta^A$ are
natural spatial coordinates.

If $N$ is considered as an independent
variable, one obtains the Einstein evolution equations with the 
super-Hamiltonian constraint as its Lagrange equation.
Allowing the lapse to depend explicitly on the spatial metric leads to new
evolution equations which differ by multiples of that constraint. (The
Hamiltonian approach widens this freedom to allow explicit dependence on
the gravitational momenta.)  For diagonal metrics, the useful lapse
choices are power law lapses, namely products of powers of the diagonal
metric components [14,15]. The Taub time gauge choice is an example.

\section{Exact Solutions}

All known exact spatially homogeneous solutions of gravitational field
equations are such that the nongravitational variables decouple from the
field equations for the gravitational variables in the following sense.
The source
energy-momentum (including a possible cosmological constant) may be 
represented in terms of the metric and constants of the motion, leading to
an entirely geometric system.  This occurs only when the source has 
either discrete or continuous additional symmetry.
For the diagonal case, the spatial curvature potential is a linear
combination of individual potential terms which are power law in the diagonal
metric components and therefore exponentials whose arguments are linear 
functions of the natural inertial  coordinates [6].
To admit exact solutions, the source can add at most more such terms to
the total potential energy $U$ when re-expressed in terms of 
possible constants of the motion.
Independent of this, one must have a suitable decoupling of the gravitational
modes for exact solutions to exist. At most one power of any exponential
argument can appear in the total potential, i.e., the exponentials with 
different arguments must be associated with linearly independent directions
in the space $\mathcal{M}\dee$.  If this is so, one may obtain integrable
decoupled equations,
the only complication being the causal nature of the level hypersurfaces
(hyperplanes) of the individual exponential potentials.
These hyperplanes may be timelike, so that level surfaces move with speed
$ds/d\alpha<1$ in $\beta^A$ space, or spacelike or null, in the latter case
moving with speed $ds/d\alpha=1$. If different powers of a given exponential
argument appear, the resulting integrals for decoupled modes become more
complicated.

Consider first the four-dimensional vacuum case (with zero cosmological
constant).  The diagonal exact solutions fall into two overlapping families
of solutions which intersect at the well known Kasner family [16].
The first is the
family found by Taub [17] in his introduction of the spatially homogeneous
cosmological models.  He used the time gauge $N=g^{1/2}$, namely the Taub
time gauge apart from a constant.
The second family is due to Joseph [18] and 
Ellis and MacCallum [19] and contains a subfamily which can be analytically
continued to the Kantowski-Sachs models [20]. These latter models are the
only nontrivial multiply transitive symmetry type
for this dimension in the sense
that they admit no subgroup acting simply transitively on the spatial sections
and so cannot be represented as a simply transitive case with higher symmetry.
The Taub family contains one or two nonnull potentials while the
Joseph-Ellis-MacCallum family contains one null potential in Taub time gauge.
(Changing the time gauge may change the causality properties.)
The Joseph-Ellis-MacCallum solutions have been given in the time gauge
$N=(g_{33})^{1/2}$ but can easily be converted to the Taub time gauge.
These two families of exact solutions are very useful examples.

The Taub family of solutions may be described by
$$\eqalign{& \C123=\C231=n^{(1)}\ , \qquad  \C312=n^{(3)}\ , \qquad
\bfg\dee=e^{2\alpha}e^{2(\beta^+\bfe_+ + \beta^-\bfe_-)}\ ,\cr &
\bfe_+=\diag(1,1,-2)\ , \qquad 
\bfe_- =\sqrt3 \diag(1,-1,0)\ ,\qquad \beta^-= \delta(n^{(1)}) \beta^-_{0}\ ,
\cr}
\eqno(13)$$
and their dynamics in Taub time gauge
is described by the Hamiltonian (expressed in velocity
phase space)
$$ H\ttg=\half (-\dot{\alpha}{}^2 +\dot{\beta}{}^{+\,2} +\dot{\beta}{}^{-\,2})
+6\nthree e^{4(\alpha-2\beta^+)} -24\none e^{-2(\beta^+-2\alpha)}\ .
\eqno(14)$$
The Lorentz transformation [4]
$$ (\alphab,\betab^+) =3^{-1/2}(2\alpha-\beta^+,-\alpha+2\beta^+)
\eqno(15)$$
leads to a completely decoupled Hamiltonian
$$\eqalign{ H\ttg
 &= \half (-\dot{\alphab}{}^2 +\dot{\betab}{}^{+\,2} +\dot{\beta}{}^{-\,2})
 +6\nthree e^{-4\sqrt3\betab{}^+} -24\none e^{2\sqrt3 \alphab}\cr
 & = -H_{\alphab} +H_{\betab{}^+} +H_{\beta^-} =0\cr}
\eqno(16)$$
consisting of three 1-dimensional scattering problems with exponential
potentials and constant energies, restricted only by the constraint that
the appropriately signed sum of the individual energies vanish.  The
solution of an exponential scattering problem leads to hyperbolic or
trigonometric functions
$$\eqalign{& H_x= \half(\dot{x}{}^2 +\mu e^{\nu x}) =E_x
\quad\rightarrow\quad t=\int dx (2E_x-\mu e^{\nu x})^{-1/2}\ ,\cr
&\qquad \hbox{or defining $\gamma=(2E_x)^{1/2}$}\ ,\cr
&e^{-{\half}\nu x}=\cases{ \mu^{1/2}
\gamma^{-1} \cosh(\half\gamma\nu t) & $\mu>0$
\ ,\qquad $E_x>0$ \cr
|\mu|^{1/2}
\gamma^{-1} \sinh(\half\gamma|\nu t|) & $\mu<0$\ ,
   \qquad $E_x\in \Re$\ .\cr} \cr}
\eqno(17)$$

$(\overline{\alpha},\overline{\beta}{}^+,\beta^-)$ are inertial coordinates
of the rest frame of the $n^{(3)\,2}$ potential, which moves with speed
$d\beta^+/d\alpha=\half$ in $\beta^A$ space.  For the Bianchi type IX case,
characterized by $n^{(3)}n^{(1)}>0$ (so $\beta^-=0$),
the hyperbolic cosine solution is relevant,
interpolating between the asymptotic free positive and negative exponential
solutions (for the metric components) at $t=\pm\infty$; the unit velocities
in $\beta^A$ space of the asymptotic solutions are related by a simple
reflection in the rest frame of this potential [4].  Letting $n^{(1)}\to0$
contracts the group to Bianchi type II, eliminating the ``tachyonic"
$n^{(3)}n^{(1)}$ potential (it moves with speed $d\beta^+/d\alpha=2$ in
$\beta^A$ space) and allows free motion parallel to the $n^{(3)\,2}$
potential (the $\overline{\alpha}$ and $\overline{\beta}{}^-$ directions).
For the Bianchi type VIII case, characterized by $n^{(3)}n^{(1)}<0$, only
the positive energy solutions are relevant for the $\overline{\alpha}$ motion
(the potential is negative) due to the super-Hamiltonian constraint.
When both $n^{(1)}$ and $n^{(3)}$ vanish, one obtains the Abelian case of
Bianchi type I and all three variables have free motion, leading to the
Kasner solution when re-expressed in proper time gauge, the exponentials in
the Taub time converting to powers of the proper time.

The Joseph-Ellis-MacCallum family is also
described by a Hamiltonian system for
a diagonal metric matrix but with structure constants and Taub time gauge
potential given by
$$\eqalign{ &\C131=a+q\ ,\quad \C232=a-q\ , a^2\equiv -hq^2\ ,\quad\ 
\lambda\equiv qa^{-1} =\pm(-h)^{-1/2}\ , \cr 
 &U\ttg= 12\cdot 6 e^{4(\beta^0+\beta^+)} (a^2+q^2/3)
 = 12\cdot 6a^2e^{4(\beta^0+\beta^+)} (1+\lambda^2/3)\ .\cr}
\eqno(18)$$
The null hyperplane level surfaces of this
potential require a different transformation of coordinates when $a\neq0$
$$\vbox{\halign{$#$\hfil&\qquad $#$\hfil\cr
 \betab^0=\beta^0+ \beta^+ &\beta^0=(1+\zeta^2/3)\betab^0 -\betab^+ +\zeta
  \betab^-/\sqrt3\cr
 \betab^+= \beta^++ \zeta\beta^-/\sqrt3 &\beta^+=
\betab^+ -\zeta\betab^-/\sqrt3
  -\zeta^2\betab^0/3\cr
 \betab^-= \beta^- -\zeta(\beta^0+\beta^+)/\sqrt3 &\beta^- =\betab^-+\zeta
  \betab^0/\sqrt3 \ ,\cr}}
\eqno(19)$$
where $\zeta\equiv [1-\delta(a)]\lambda$,
leading to
$$\eqalign{
& H\ttg= -\half(1+\zeta^2/3)(\betab^0)\dt^2 +\half (\betab^0)\dt (\betab^+)\dt
 +\half(\betab^-)\dt^2 + 12\cdot 6a^2e^{4\betab^0} (1+\lambda^2/3)\cr
& \mathcal{H}^{\bot}{}_3 =-12ae^{2\betab^0}(\betab^+)\dt \qquad
 [\hbox{note:\  }a\betab^+ =a\beta^+ +q\beta^-/\sqrt3 ]\ .\cr}
\eqno(20)$$

If $a\neq0$, then $(\betab^+)\dt=0$ (set its constant value
equal to zero since it is gauge) and the Hamiltonian becomes
$$ H\ttg= -(1+\lambda^2/3)H_{\betab^0} +H_{\betab^-}\ ,
\eqno(21)$$
thus decoupling into free motion for $\betab^-$ with constant energy
(linear behavior)
and a one-dimensional scattering problem for $\betab^0$
in a negative exponential potential
with constant energy (the ``$\ln\sinh$" solution) which must be positive
to satisfy the vacuum super-Hamiltonian constraint of vanishing total energy.

If $a=0$ and therefore $\zeta=0$, then $\beta^+$ and $\beta^-$ are unchanged
by the transformation while $\betab^0=\beta^0+\beta^+$ and
it is convenient to introduce another null coordinate $\betab^+
\equiv \beta^+-\beta^0$ so that the kinetic energy becomes
$$ T\ttg=\half(\betab^0)\dt (\betab^+)\dt +\half (\beta^-)\dt^2\ .
\eqno(22)$$
Here the null  geometry is essential. $\beta^-$ and
$\betab^0$ undergo free motion
while $(\betab^+)\ddt=8 U\ttg $ leading to exponential behavior,
describing the Bianchi type VI$_0$.
However, if $q\neq0$, then the supermomentum constraint requires that
$\beta^-$ be a constant (set it equal to zero since it is gauge)
while if $q=0$ one arrives again at the Abelian case and all three variables
are free. The Bianchi type VI$_0$ solution in proper time time gauge
was given by Ellis and MacCallum [19].

The condition $a=q\neq0$ (or $a=-q$) describes Bianchi type 
$\hbox{III} =\hbox{VI}_{-1}$. The vacuum supermomentum constraint then imposes
the condition $(g_{11}/g_{33})\dt=0$ associated with local rotational
symmetry, the spatial
geometry being that of a family of geodesically parallel
2-dimensional surfaces of constant negative curvature.
Allowing $q^2$ to take negative values in the Einstein equations 
simply changes the sign of the potential and
describes
the Kantowski-Sachs models which instead have constant positive curvature
on the geodesically parallel family of 2-dimensional surfaces, changing
the solutions of the one-dimensional scattering problem from hyperbolic
to trigonometric functions.

All of these solutions describe the effect of zero, one or two independent
exponential potentials on the flat diagonal configuration space $\mathcal{M}\dee$.
They can be extended to certain nonvacuum nonzero cosmological constant cases
in four and higher dimensions in a trivial way by a mechanism called ``variation
of parameters" [15].  The constraints impose certain values on the parameters
which appear in the evolution equations.  Allowing the parameters to vary from
their constrained values is equivalent to introducing a nonzero energy-momentum
tensor which may be interpreted as that of a stiff perfect fluid moving
orthogonal to the spatial sections if the supermomentum is zero, or equivalently
as a massless scalar field.  Nonzero values of the supermomentum in the
Joseph-Ellis-MacCallum case lead to tilted fluid flow and can also be 
interpreted in terms of changing the spacetime symmetry from isometries to
homothetic transformations, leading to spatially self-similar solutions.  

The massless scalar field can be obtained by dimensional reduction from a
5-dimension\-al vacuum Kaluza-Klein theory, together with a Weyl transformation,
namely a conformal transformation of the four-dimensional metric when imbedded
in five dimensions [21].  This allows one to relate the above solutions
to solutions of the Brans-Dicke theory [21--23].
One may also extend the imbedding to vacuum Kaluza-Klein theories
with any number of additional flat dimensions, thus obtaining exact solutions
of those theories [7,24].
The Weyl transformation is related to a redefinition
of the variable $\alpha$ when adding extra dimensions.

Nonvacuum Kaluza-Klein theories also admit such an extension of the
four-dimension\-al vacuum solutions by variation of parameters. Both ten and
eleven-dimensional supergravity when restricted to the Bose sector
merely add additional exponential potentials to the spatial curvature
potentials when one assumes a Freund-Rubin [25] or Freund-Rubin-Englert [26]
ansatz,
since the differential form fields can then be expressed in terms
of the metric and constants of the motion. The scalar field in the 
ten-dimensional case is related by dimensional reduction to an extra flat
dimension [27]. One thus sees the same familiar exact solutions 
reappearing [28--31].

These solutions may also be deformed by a second kind of variation of
parameters in which one changes the parameters which appear in the
field equations themselves in such a way that the equations still remain
integrable.  In the Abelian case, the addition of a cosmological constant
or a perfect fluid with equation of state $p=(\gamma-1)\rho$ whose flow is
orthogonal to the spatial sections adds a single $\alpha$-dependent
exponential to the Hamiltonian, a timelike potential like one of the two
in the semisimple Taub case in four dimensions.
One can add such a potential to the nonsemisimple Taub case or the
Joseph-Ellis-MacCallum case and still obtain the same kinds of solutions.
On the other hand, adding a locally rotationally symmetric electromagnetic
field changes the nature of the solutions, but only deforms them in a
simple way, leading to more very similar exact solutions.
One may in fact add many fields simultaneously, leading to,
for example, solutions
in four dimensions with a stiff perfect fluid, a scalar field and an
electromagnetic field, the scalar field allowing the optional reinterpretation
in terms of solutions of the Brans-Dicke theory.

Consider adding a locally rotational electromagnetic field to the
Taub family of models in four dimensions [32].  This simply adds another
$\betab^+$ potential to the Hamiltonian but with a different power, leading
to a one-dimensional scattering problem with two different exponential
potentials.  The explicit integral in (17), when rewritten in terms of
the variable $u=\mu^{-1/2}e^{-\half\mu x}$, becomes proportional to the
standard integral ( $\int du (c^2 u^2-1)^{-1/2}$) for the inverse hyperbolic
cosine when $\mu$ and $E_x$ are positive as occurs in this case.
The additional potential merely adds a linear term to the radical which
can be eliminated by completing the square.  The new solution for $u$ is then
a constant plus another constant times the old solution, a simple
variation of parameters [15,33].

The Taub time gauge is clearly very powerful, but other power law lapse
time gauges are also useful.  Changing the lapse leads to different
decoupling possibilities for different kinds of driving terms.  For example,
adding a cosmological constant to the semisimple Taub case (Bianchi types
VIII and IX) leads to three
exponential potentials along distinct directions in the two-dimensional
Taublike subspace of $\mathcal{M}\dee$ associated with local rotational symmetry.
In addition to the orthogonal pair of
timelike and spacelike potentials, a
cosmological constant term adds a null potential,
causing a coupling of the equations.  However, in Misner's [15,34] time
gauge $N\sim (g_{33})^{-1/2}$, one degree of freedom remains unchanged
but enters
the equation of motion for the second in such a simple way that the new 
solution merely adds on a term to the old one [35].
There are many other instances in various time gauges where similar ideas
apply (including the Joseph-Ellis-MacCallum family in the original time gauge
and even the familiar Friedmann models in conformal time gauge 
$N\sim e^{\alpha}$, proper time time gauge  and other time gauges, as well
as higher dimensional generalizations of these solutions).
However,
no systematic analysis of lapse choices has ever been undertaken as a way
of simplifying the equations of motion, other than the limited discussions
of references [7] and [15].

It  is incredible how many pages of journal space have regurgitated the same
old solutions in various disguised forms without leading to some recognition
of the connections. Too many people interested in exact cosmological
solutions suffer to some degree from a sort of tunnel vision which prevents
them from taking the time to appreciate the ADM perspective for the very
branch of gravitation for which it is the most powerful.
The present article shows that the rewards for doing this are considerable.

\section{Acknowledgment}

J. Demaret is thanked for providing the stimulus to consider these questions
and L. Angelini for helpful discussion during the  course of their development.

\end{document}